\documentclass{JHEP3}
\usepackage{amsfonts}
\usepackage{amsmath}
\usepackage{amssymb}
\usepackage{epsfig}
\usepackage{graphicx}
\newcommand{\bea}{\begin{eqnarray}}
\newcommand{\eea}{\end{eqnarray}}

\newcommand{\dd}{\mbox{d}}

\preprint{
SNUST 060601 \\ UT-06-10 \\
APCTP 2006-002\\
 {\tt hep-th/0606127} } \vskip1cm
\title{The Nothing at the Beginning of the Universe\\
Made Precise}
\author{Yu Nakayama${}^{a}$, Soo-Jong Rey${}^{b}$, Yuji Sugawara${}^{a}$
 ~~~~~~~~~~~\\
 ~~~~~~~~~~~\\
{\sl ${}^{a}$ Department of Physics, Faculty of Science,
University of Tokyo\\
7-3-1 Hongo, Bunkyo-ku, Tokyo 113-0033  {\rm JAPAN}\\
${}^{b}$ School of Physics and Astronomy \& F.P.R.D. \\
Seoul National University, Seoul 151-747 {\rm KOREA}}
~~~~~~~~~~~~~~~~\\
\email{ nakayama, sugawara@hep-th.phys.s.u-tokyo.ac.jp, \hskip0.5cm
sjrey@snu.ac.kr} }
\abstract{ We propose a new worldsheet approach to the
McGreevy-Silverstein proposal: resolution of spacelike singularity
via Scherk-Schwarz compactification and winding string condensation
therein. Our proposal is built upon so-called three parameter
sine-Liouville theory, which has useful features and could be
solvable in conformal field theory method. Utilizing standard Wick
rotation, we compute string pair production rate exactly in terms of
renormalized worldsheet cosmological constant and find that the
production rate is finite for six or less spacetime dimensions. We
also find that the sine-Liouville potential excises string
excitation in the asymptotic past, and that such "Nothing state" is
realizable for a range of sine-Liouville coupling constants. We
compute one loop vacuum-to-vacuum transition amplitude and again
detect presence of the "Nothing state". We also survey various
worldsheet approaches to the tachyon condensation based on timelike
Liouville theory. We point out that string theory on a conifold
provides the upper critical dimension for realizing the "Nothing
state", thus making contact with the blackhole / string transition
point.}
\begin{document}
\section{Introduction}
According to the standard 'big-bang' cosmology, the universe has
started from a spacelike singularity. Within general relativity, it
is established that such singularity is unavoidable. Near the
singularity, curvature of the spacetime blows up, and classical
treatment of the gravity breaks down. An outstanding question is
whether effects beyond classical gravity can resolve the big-bang
singularity. In the context of string theory, such effects may
originate from string worldsheet and quantum fluctuations. As such,
variety of possible string theoretic mechanisms for resolving
spacelike singularity have been proposed in the past \cite{martinec}
-- \cite{easson}.

Recently, McGreevy and Silverstein (MS)
proposed an extremely interesting mechanism
for resolving cosmological singularity \cite{MS}.
Their proposal relies on string theory;
central to their proposal is utilization
of "tachyon condensation" of winding string
around cosmologically shrinking $\mathbb{S}^1$,
which in turn creates a mass gap to
closed string worldsheet degrees of freedom \cite{polyakov}.
Take, for example,
Type II superstring in an expanding universe of topology
 $\mathbb{R}_t \times \mathbb{S}^1_\theta \times {\cal M}_\perp$:
\bea \dd s^2_{\rm II} = \ell_{\rm st}^2 [ - \dd t^2 + a^2(t) \dd
\theta^2 + \dd s_\perp^2  ], \, \label{background} \eea
where the scale factor $a(t) = a_o t^\nu$ is driven by a homogeneous
and isotropic energy-momentum tensor. Around $\mathbb{S}^1$, one
assigns spacetime fermions to obey `thermal' boundary condition
\cite{atickwitten}, thus cosmologically realizing the Scherk-Schwarz
compactification \cite{scherkschwarz}. Accordingly, the spacetime
supersymmetry is broken, causing Bose-Fermi mass splitting set by
$a(t)/ \ell_{\rm st}$. In particular, winding string state around
the Scherk-Schwarz circle $\mathbb{S}^1(\theta)$ survives the
twisted Gliozzi-Scherk-Olive (GSO) projection \cite{GSO} and has
mass spectrum $ m^2_{w} = (a^2(t) -1)/ \ell_{\rm st}^{2}$.
Near the cosmological singularity
at $t=0$, proper size of the $\mathbb{S}^1$ shrinks smaller than the
string scale $\ell_{\rm st}$. The winding state becomes tachyonic,
so it condenses over a region of the spacetime near the cosmological
singularity. By taking adiabatic regime $0 < \nu \ll 1$, one can
arrange the tachyon rolling to take place while string coupling and
Hubble parameter remain small~\footnote {Winding tachyon
condensations was investigated {\it e.g.} in \cite{BarRab} --
\cite{HikTai}
in relation to various dynamical aspects
of superstring theory, say, removing singularities,
topology changes or space-time phase transitions, etc.
Previous studies
of cosmological backgrounds
based on analytically continued solvable conformal field theories
with linear dilatons are given {\it e.g.} in
\cite{EGKR} -- \cite{TT}. See also \cite{Craps} for a review
and more complete reference list.}.
Dynamics of a probe string is then describable by worldsheet theory
involving Lorentzian signature ($\mathcal{N}=1$ supersymmetric)
sine-Liouville theory \cite{FZZ2,fateev,Kim}
~\footnote{ We suppress
decoupled free conformal field theory for transverse space in
(\ref{background}) and
Faddeev-Popov ghosts. We also suppress the fermionic degrees
of freedom.}
\bea {\cal Z}_{t, \theta} \simeq \int [\dd t \dd \theta ] \, \exp
\left( {i \over \ell_{\rm st}^2} \int \dd^2 \sigma \big[ - |\partial
t|^2 + |\partial \theta|^2 - \mu^2 e^{- 2 \kappa t} \cos^2 (\omega
\widetilde{\theta}) +\cdots \big] \right) \,
\label{lorentzian} \eea
where ellipses abbreviate terms involving worldsheet fermions. MS
examined string dynamics in this region and concluded that, if $\mu
$ is real (which plays the role of worldsheet cosmological
constant), mass gap is generated for all worldsheet fields. In
spacetime picture, this means that all closed string excitations are
lifted up infinitely heavy. In effect, the tachyon condensation is
seen to excise the epoch of big-bang singularity and replacing it by
so-called "Nothing state". What makes MS proposal particularly
attractive is that the "Nothing state" is Lorentzian, string
theoretic realization of the Hartle-Hawking no boundary proposal
\cite{hartlehawking}. MS also argued that their proposed mechanism
is generic and can be extended to different types of spatial
topology and boundary condition.

In drawing their claims from string worldsheet analysis, MS relied
largely on saddle-point approximation. Such approximation is not
likely to be self-consistent. Given that the singularity excision
takes place within a string scale sized region of the spacetime
(\ref{background}), full-fledged worldsheet treatment of both the
tachyon condensation and the string propagation therein are
indispensable. Within the setup of MS proposal, however, such
treatment is extremely difficult if not impossible~\footnote{At
best, (\ref{ms}) may be treated as a variant of $A_2$ Toda field
theory.}. This is because, for the Euclidean signature
($\mathcal{N}=1$ supersymmetric) sinh-Liouville theory
\bea S_{\rm MS} = {1 \over 2 \pi} \int \dd^2 z \, \big[ \partial
\varphi \overline{\partial} \varphi + \partial \theta
\overline{\partial} \theta + 2 \pi \mu^2_{\rm E} e^{ - 2 \kappa
\varphi} \cosh^2 (\omega \widetilde \theta) + \cdots \big]
\label{ms} \eea
adopted by MS in prescribing the theory (\ref{lorentzian}),
systematic conformal field theory treatment is currently
unavailable. The negative coupling of Wick-rotated tachyon
background term ($\mu \rightarrow i \mu_{\rm E}$) cause the theory
non-unitary and renders the difficulty even worse.

Given such difficulties, a host of physics questions arise. Is the
tachyon condensation, especially dependence of physical processes to
the cosmological constant, describable by exact conformal field
theory approach? After tachyon condensation, is the spacetime
geometry deformed only locally near the singularity or over an
extended region? Is the rate of Bogoliubov pair production, which is
caused by time-dependent background, finite and controllable? Is the
"Nothing state" replacing the singularity unique or are there many
possibilities? Related to it, given that the cosmological constant
$\mu$ is subject to renormalization, is its phase determinable after
string worldsheet effects are fully taken into account?

In this work, we propose a new worldsheet approach
to winding tachyon condensation which bypasses
all aforementioned difficulties.
It is based on so-called
the three-parameter model
of the Euclidean signature sine-Liouville theory:
\bea
S_{\rm NRS} = {1 \over 2 \pi} \int \dd^2 z \,
\big[ \partial \varphi \overline{\partial} \varphi + \partial \phi_1 \overline{\partial} \phi_1 + \partial \phi_2 \overline{\partial} \phi_2 + 2 \pi \mu \, e^{\alpha \varphi}
\cos (\beta \phi_1 + \delta \phi_2) + \cdots \Big] \ . \label{3paramodel} \eea
%
This theory is treatable by standard conformal field theory
approach, in sharp contrast to the sinh-Liouville model (\ref{ms}).
Consequently, we can examine full-fledged stringy effects to
excision mechanism of the cosmological singularity and to onset and
evolution of the tachyon condensation. Of particular interest is
whether the `Nothing state' is robust enough not to be affected by
string worldsheet effects. We shall be able to answer these
questions affirmatively by extracting renormalization effect to the
worldsheet cosmological constant $\mu$ in a precise manner. The most
significant point is that our results go beyond semiclassical
approximations; we were able to do so since the proposed worldsheet
theory (\ref{3paramodel}) is treatable in full-fledged conformal
field theory approach.


In section 2, we first recapitulate the three-parameter
sine-Liouville theory. We then propose a prescription of Lorentzian
sine-Liouville theory relevant for the winding string tachyon
condensation via the standard Wick rotation of the Euclidean theory.
In section 3, via the Wick rotation, we propose a concrete
realization of the "Nothing state" in a precise manner. Using
conformal field theory approach, we extract bulk reflection
amplitudes of the sine-Liouville theory and, from this, obtain the
Bogoliubov pair production amplitudes by performing the Wick
rotation carefully. We study renormalization of the tachyon mass
parameter (worldsheet cosmological constant) and find that the
`Nothing state' indeed emerges over a range of worldsheet
sine-Liouville coupling parameters. We also study the one-loop
vacuum-to-vacuum transition amplitude, again via the standard Wick
rotation, and find that it leads to the same conclusion as those
based on the Bogoliubov amplitudes. In section 4, we first discuss
several possible Liouville theory models of winding tachyon
condensation for various worldsheet supersymmetries. We then show
that realistic tachyon condensation with ${\cal N}=1$ superconformal
symmetry is mappable to the three-parameter sine-Liouville theory.
We also point out that the upper critical dimension for realizing
the "Nothing state" corresponds to string theory on a conifold and
that this coincides with the black hole -string transition point. We
finally present an intuitive spacetime picture that underlies
emergence of the "Nothing state".

\section{Proposal}
Begin with our proposal for the cosmological winding tachyon
condensation. A criterion we set out for a viable worldsheet
approach is that the standard Wick rotation  of cosmological string
background is achievable within well-defined conformal field
theories. Though the setup also adopts specifically cosmological
Scherk-Schwarz compactification and tachyon condensation of winding
string near the cosmological singularity, our proposal differs
significantly from MS proposal. First, our proposal is based on
well-defined conformal field theory at the outset: the cosmological
background is described by first starting from Euclidean signature
sine-Liouville (instead of {\sl sinh-Liouville}) theory and then
make the standard Wick rotation to Lorentzian signature
sine-Liouville theory. Second,  utilizing conformal field theory
approach, our treatment is capable of capturing full-fledged
worldsheet effects describing spacetime dynamics all the way down to
string scale.

\subsection{Three-parameter sine-Liouville theory}
Our proposal is based on so-called three-parameter sine-Liouville
field theory \cite{Kim}, which is a generalization of the models
with smaller number of parameters \cite{FZZ2, fateev}. This is a
theory defined in terms of three spacelike scalar fields: a
Liouville field $\varphi$ and two compact scalar fields $\phi_1,
\phi_2$, and is described by the action~\footnote{We adopt
conventions that set $\dd^2 z \dd \sigma^1 \dd \sigma^2$,
$\ell^2_{\rm st} = 2\pi$ so that $\langle
\varphi(z,\bar{z})\varphi(w, \bar{z}) \rangle = -\log \vert z - w
\vert^2 $, etc. }
\begin{eqnarray}
S = \frac{1}{2\pi}\int \dd^2z \left(\partial\varphi
\bar{\partial}\varphi +
\partial\phi_1\bar{\partial}\phi_1 +
\partial\phi_2\bar{\partial}\phi_2 + 2\pi \mu \,
e^{\alpha\varphi}\cos(\beta\phi_1+\delta\phi_2) + \frac{q}{2}
{\cal R}_{(2)} \varphi \right) \label{3parasl}
\end{eqnarray}
where the coupling parameters $\alpha, \beta, \delta$ and the
background charge $q$ are all real-valued. The last two terms
describe sine-Liouville tachyon and linear dilaton backgrounds. For
the moment, we view (\ref{3parasl}) as a model exhibiting the
`Nothing state' at the beginning of the universe. In this context,
we view $\phi_1$ as the T-dual variable of $\widetilde{\theta}$ in
\eqref{lorentzian} and $\phi_2$ as other relevant $c=1$ worldsheet
degrees of freedom.  In section 4, we shall show that the proposed
model can be embedded into Type II string theory once we set
$\delta=1$ and fermionizing $\phi_2$.

The theory (\ref{3parasl}) is characterized by the coupling
parameters $(\alpha, \beta, \delta)$.  The background charge $q$ is
related by the conformality condition to these parameters as
\begin{eqnarray}
q = {1 \over 2 \alpha} (\alpha^2 -\beta^2-\delta^2 +2)
\label{confcond}
\end{eqnarray}
and sets total central charge of the theory to $c = \overline{c} = 3
+ 12q^2$. To ensure that the sine-Liouville potential cuts off the
strong coupling region, we shall restrict to the range $q \alpha
> 0$. The central charge is invariant under the exchange map
\bea \alpha \longleftrightarrow \widetilde{\alpha} \equiv {1 \over
\alpha} (2 - \beta^2 - \delta^2),  \eea
but, unlike the Liouville theory, this map does not yield
self-duality of the theory. For example, the correlators are not self-dual but are
paired up between $\alpha$ and $\widetilde{\alpha}$ theories.
A special subset of the theory is the two-parameter family $2\alpha
q=1$ \cite{fateev} with $\delta = 1$ (i.e.
$\alpha=\beta=\frac{1}{2q}$), the well-known ${\cal N}=2$ Liouville
theory, which is mirror to super-coset SL$(2,\mathbb{R})_k/$U(1)
theory \cite{FZZ2,GK,HoriKap,Tong}. In this case, the $\phi_2$ boson
arises from bosonizing the ${\cal N}=2$ superconformal fermions.
Finally, the $c=3$ sine-Liouville theory is obtainable by taking the
limit $q \rightarrow 0$.

In the asymptotic region $\varphi \rightarrow -\infty$, the tachyon
condensate vanishes.
The theory reveals
U(1)$\times\widetilde{\rm
U}(1)$ worldsheet symmetries, whose holomorphic currents are
\bea J = i \Big({1\over \beta} \partial \phi_1 -{1 \over \delta}
\partial \phi_2\Big); \qquad \widetilde{J} =
i {\beta
\partial \phi_1 + \delta \partial \phi_2 \over \beta^2 + \delta^2},
\nonumber \eea
normalized so that the tachyon condensates carry integer-valued
charges. For primary vertex operators, conformal dimension and
global charges are given by
\bea V(a,b,c) = e^{a \varphi + ib \phi_1 + ic \phi_2}: \quad
\Delta_V &=& \frac{1}{2}(-a^2 + 2 a q + b^2 + c^2) \nonumber \\
Q_V &=& {b \over \beta} - {c \over \delta} \nonumber \\
\widetilde{Q}_V &=& {b \beta + c \delta \over \beta^2 + \delta^2}.
\nonumber
\eea
The tachyon background in (\ref{3parasl}) is neutral under U(1) but
carries $\pm 1$ unit charges under $\widetilde{\rm U}(1)$. As such,
at finite $\varphi$, the tachyon condensate breaks $\widetilde{\rm
U}(1)$ symmetry to $\mathbb{Z}$. Hereafter, we shall refer U(1)
charge the momentum quantum number and $\widetilde{\rm U}(1)$ charge
as the winding number of the string around the Scherk-Schwarz
compactification circle.

For later considerations, we are interested in two-point correlation
functions. By examining U(1), $\widetilde{\rm U}(1)$ quantum
numbers, we see that there are two possible classes of two-point
correlation functions~\footnote{Our conventions for the
SL$(2,\mathbb{C})$ states, charge conjugation and normalization are
such that
\bea \Big< V(2q-a, b, c) V(a, -b, -c) \Big> = 1. \nonumber \eea }:
\bea {\cal R} \equiv \Big< V(a,b,c) V(a, -b, -c) \Big> \qquad
\mbox{and} \qquad
{\cal D} \equiv \Big< V(a,b,c) V(a,b,c) \Big>. \nonumber \eea
They define two possible types of reflection amplitudes: as the
incoming wave hits the sine-Liouville potential, part of the
reflection amplitude conserving the winding quantum number ($\Delta
\widetilde{Q} = 0$) is described by ${\cal R}$, while the part
shifting the winding number   by integer unit ($\Delta \widetilde{Q}
= \pm1, \pm2, \dots$) is described by  ${\cal D}$. Thus, compared to
the Liouville theory, the correlators ${\cal D}$ are new and highly
nontrivial. Based on recursion relations for different $\alpha$,
results for integer-valued $a,b,c$ and analytic continuation
thereof, a closed form of ${\cal D}$ was proposed in \cite{Kim}.
As the simplest situation, consider scattering a mode carrying
vanishing momentum and winding quantum numbers off the
sine-Liouville potential. The reflection amplitude is given by
\begin{eqnarray}
&& {\cal D} (a,0,0)\Big\vert_{Q_V =0} = {\cal R} (a,0,0) \nonumber \\
&=& (\xi)^{\frac{q-a}{\alpha}}
\frac{\Gamma(1+\frac{a-q}{\alpha})\Gamma(\frac{q}{\alpha}
-\frac{a-q}{\alpha})\Gamma(1+\alpha(a-q))\Gamma(1-\alpha(a-q)+\alpha
q)}{\Gamma(1-\frac{a-q}{\alpha})\Gamma(\frac{q}{a}+\frac{a-q}{\alpha})
\Gamma(1-\alpha(a-q))\Gamma(1+\alpha(a-q)+\alpha
q)} \nonumber \\
&& \nonumber \\
&\times& \exp\left[\int_0^\infty \frac{{\rm d}
s}{s}\left(\frac{e^{-(1+\alpha^2+\alpha a - 2\alpha q)s} -
e^{-(1+\alpha^2 -\alpha a)s}}{(1-e^{-s})(1-e^{-\alpha^2 s})}
(1-e^{-2\alpha q s}) + 4(a-q)qe^{-s}\right)\right] \label{D} .
\end{eqnarray}
Here, $\xi$ denotes the "renormalized" cosmological constant on the
worldsheet:
\begin{eqnarray}
\xi = (\pi\mu)^2 \frac{{\gamma}(-(\alpha^2+\beta^2+\delta^2))}
{\gamma(-\frac{1}{2}(\alpha^2+\beta^2+\delta^2))} \gamma(1+\alpha^2)
\gamma(1+\frac{1}{2}(\alpha^2-\beta^2-\delta^2)) \ , \label{xi}
\end{eqnarray}
where $\gamma(x) \equiv \Gamma(x)/\Gamma(1-x)$~\footnote{We are
using the different notation ($\alpha'=2$) from that in \cite{Kim}
($\alpha'=1$). We have also corrected typos in the proposal
\eqref{D}, which was confirmed by the authors of \cite{Kim}.}. The
last line originates from solving the recursion relations in terms
of the integral representation of the digamma function.

Classical limit of the three-parameter sine-Liouville theory
(\ref{3parasl}) should be well-defined. We see from (\ref{3parasl})
that $\varphi, \phi_1, \phi_2$ dynamics becomes semiclassical in the
limits $\alpha \rightarrow 0, \beta \rightarrow 0, \gamma
\rightarrow 0$, respectively~\footnote{Note that the validity of the
minisuperspace analysis also forces us to take $\beta \to 0$ limit
so as to treat $\phi_1$ as a weakly interacting field. Without
taking the adiabatic limit for $\phi_1, \phi_2$, the minisuperspace
analysis may not make sense due to the unboundedness of the
sine-Liouville potential. We, nevertheless, believe that the theory
is well-defined as a quantum theory. For example if one fermionizes
$\phi_2$ at a special radius as we will do later, the potential term
turns out to be a time-dependent mass term for worldsheet fermion.
This is certainly a stable interaction classically. We would like to
thank H.~Ooguri for pointing out the unboundedness of the
sine-Liouville potential.}.  Because of the conformality condition
(\ref{confcond}), simultaneous classical limit is achieved if the
dilaton slope $q$ is set to $1/\alpha \rightarrow \infty$. In these
classical limits, the renormalized cosmological constant $\xi$ is
replaced by a classical cosmological constant, which is always
proportional to $\mu^2$. The unitarity condition, $|{\cal
R}(a,0,0)|=1$, restricts the coupling parameters further. For
real-valued momentum $p$ where $ip = a-q$, it is straightforward to
check that the ratios of gamma functions in (\ref{D}) are pure
phases. The third line in (\ref{D}) is also a pure phase in so far
as $\alpha^2 + \beta^2 + \delta^2 >0$, which is always guaranteed in
Euclidean theory (but will be nontrivial in Lorentzian theory). The
factor $(\xi)^{(q-a)/\alpha}$ is also a pure phase {\sl provided} we
restrict the coupling parameters in the neighborhood $\alpha, \beta,
\gamma \rightarrow 0$ to range over the domain $\alpha^2 > \beta^2 +
\delta^2$.

Once we set $2q\alpha = 1$, the proposed reflection amplitude is
consistent with that for the two-parameter Sine-Liouville theory
\cite{fateev}. On the other hand, if one sets $\beta = \delta = 0$,
the reflection amplitude reduces to that of the pure
($\mathcal{N}=0$) Liouville theory \cite{Zamolodchikov:1995aa}. In
this case,
 we verified that the reflection amplitude \eqref{D} satisfies not
 only the ordinary shift relation:
\begin{eqnarray}
{\cal D} (a+\alpha,0,0) = {\cal D} (a,0,0) \, C^{-1}_{-}(a;\alpha)
\end{eqnarray}
but also the dual shift relation,
\begin{eqnarray}
{\cal D} (a+\alpha^{-1},0,0) = {\cal D} (a,0,0) \,
\widetilde{C}^{-1}_{-}(a;\alpha),
\end{eqnarray}
which is highly nontrivial given that\cite{Kim} did not rely on any
information of the dual recursion relations. Here, $C_-(a; \alpha)$
and $\widetilde{C}_-(a;\alpha)$ refer to the structure constants
\cite{FZZ}.

A comment is in order. In \cite{Kim}, the reflection amplitude
(\ref{D}) was deduced from extrapolation of the amplitude at integer
values of the Liouville momenta $a, b, c$. It is possible that the
extrapolation misses a possible extra contribution, which vanishes
at those discrete momenta. On the other hand, we have seen above
that (\ref{D}) passes consistency checks when reduced to Liouville
theory. As such, we feel that the expression (\ref{xi}) of the
renormalized cosmological constant is actually the correct result,
valid for all values of the Liouville momenta. Throughout this work,
we shall proceed with such tacit assumption.

\subsection{Wick rotation to Lorentzian target space}
String propagation in the Lorentzian cosmological background
(\ref{background}) can now be described via the Euclidean signature
sine-Liouville theory (\ref{3parasl}). This requires a suitable
prescription for analytically continuation between the Euclidean and
the Lorentzian target spaces \footnote{At the same time, we need to
rotate the worldsheet from Euclidean to Lorentzian one with the
standard $+i\varepsilon$ prescription for the Feynman boundary
condition. Notice that the conformal invariance remains intact
throughout these analytic continuations.}. The prescription is
expected to depend on physical situations, but should not be
arbitrary. In particular, any physically sensible analytic
continuation ought to be compatible with unitarity, causality and
analyticity. Here, for the worldsheet theory (\ref{3parasl}), we
argue that analytic continuation via the standard Wick rotation:
\begin{eqnarray}
\varphi & \rightarrow & e^{i \epsilon} t \cr \alpha &\rightarrow &
e^{i \epsilon} \kappa \cr q &\rightarrow & e^{i \epsilon} Q
\qquad\qquad \mbox{where} \quad\qquad \epsilon \rightarrow +\pi/2
\cr \phi_{1,2} &\rightarrow& \phi_{1,2} \cr \mu &\rightarrow& \mu
\label{analy}
\end{eqnarray}
satisfies all these requirements. Range of the coupling constants
$\beta, \delta$ is then fixed by Wick rotation of the conformality
condition (\ref{confcond}):
\begin{eqnarray}
2\kappa Q = \kappa^2 + \beta^2 + \delta^2 - 2 \ .
\label{Lconfcond}
\end{eqnarray}
Analytic continuation for the parameters $\alpha, q$ is the same as
that for the variable $\varphi$, and this is exactly the same as the
standard Wick rotation for a canonically conjugate pair \footnote{
Similar analytic continuation was advocated previously in the
context of the Liouville theory \cite{gutperlestrominger,stromingertakayanagi}.} The
worldsheet action is now given as
\begin{eqnarray}
S_{\rm L} = \!\!\int \!{{\rm d}^2 \sigma \over 2 \pi} \Big[\!\!-\partial_\mu t
{\partial}^\mu t +
\partial_\mu \phi_1 {\partial}^\mu \phi_1 +
\partial_\mu \phi_2 {\partial}^\mu \phi_2 + 2 \pi \mu e^{- \kappa
t}\cos(\beta\phi_1 + \delta\phi_2) - {Q\over2} {\cal R}_{(2)} t
\Big]. \,\,\,\,\,\, \label{lorentzS}
\end{eqnarray}
Since the conformal invariance remains intact throughout the
analytic continuation (\ref{analy}), the Lorentzian sine-Liouville
theory (\ref{lorentzS}) is a well-defined conformal field theory.
The worldsheet action (\ref{lorentzS}) now describes string
propagation on the Lorentzian cosmological background
(\ref{background}) with the Scherk-Schwarz compactification.
More precisely, as mentioned already, to obtain the superconformal
model relevant for heterotic or Type II superstring theories, we
need to set $\delta = 1$ and fermionize $\phi_2$ back to worldsheet
fermions.

Notice that in our proposed analytic continuation (\ref{analy}) the
worldsheet cosmological constant $\mu$ remains intact. Thus, the
generating functional
\begin{eqnarray}
{\cal Z}[\mu] := \int_{{\cal C}} [\dd t \dd \phi_1 \dd \phi_2] \,
\exp ( i S_{\rm L}) \label{partitionfunction}
\end{eqnarray}
is an unambiguous and well-prescribed function of $\mu$ irrespective
of the signature of the target spacetime, equivalently, the contour
${\cal C}$ prescribed in defining the worldsheet functional integral
(\ref{partitionfunction}). As such, one can evaluate the generating
functional utilizing the method of \cite{wise}, viz. by first
computing $\partial {\cal Z}[\mu]/\partial \mu$ and then integrating
back with respect to $\mu$ with the boundary condition that ${\cal
Z}[\mu=0]$ yields the worldsheet partition function for $c=3$ free
field theory.

We again emphasize that the crucial advantage of our proposed
worldsheet model is that the Euclidean theory (\ref{3parasl}) is a
well-defined, unitary conformal field theory for {\sl any} value of
the linear dilaton and that the standard Wick rotation (\ref{analy})
manifestly carries over conformal properties to the Lorentzian
theory.

\section{The "Nothing state" made precise}
In order for the proposed Lorentzian three-parameter sine-Liouville
theory (\ref{lorentzS}) to realize a technically natural "Nothing
state" possessing predictability and calculability, it should be
that
\begin{itemize}
\item back-reaction to the background is controllably small,
\item no string excitation is present for $t \rightarrow -
\infty$.
\end{itemize}
To ensure the first condition, which provides technical naturalness,
string coupling parameter $g_{\rm st} = e^{-Q t/2}$ should be small
outside the "Nothing state". To achieve this, we restrict $Q$ to the
range $Q \ge 0$. In case $Q
>0$, we can then confine the strong string coupling region within
the "Nothing state". In case $Q = 0$, we can keep the string
coupling fixed and small by adjusting the coupling constants $\beta,
\delta$ appropriately so that the conformality condition
(\ref{Lconfcond}) is satisfied.

It now remains to ensure that string worldsheet effects are small
enough. This is nontrivial since, even for free string theory,
time-dependent background triggers Bogoliubov pair production. Thus,
to ensure the first condition, we need to restrict the coupling
constants $\kappa, \beta, \delta$ so that the pair production rate
is sufficiently suppressed exceeding the Hagedorn growth of state
density. This is a quite non-trivial issue in which the worldsheet
quantum effect could play essential roles. In this section, we
compute the rate and show that the condition is met for a
non-trivial range of the coupling constants.

To ensure the second condition, which defines the "Nothing state",
we need to set $\kappa
> 0$ and to let the sine-Liouville potential provides an impenetrable
barrier to closed string excitations in the asymptotic past $t
\rightarrow - \infty$. That such a barrier is present, however,
cannot be seen within the minisuperspace approximation. For one
thing, the sine-Liouville potential is unbounded from below for a
range of the compact boson fields, where $\cos (\beta \phi_1 +
\delta \phi_2)$ takes a negative value. To probe the presence of
potential barrier and hence the putative "Nothing state", we shall
compute one-loop vacuum transition amplitude in section 3.3 and
examine its extensivity with the temporal volume.

\subsection{Bogoliubov pair production}
Now, string dynamics (\ref{lorentzS}) in the  cosmological
background (\ref{background}) would involve Bogoliubov string pair
production where the winding tachyon rolls rapidly during the early
epoch of the big bang~\footnote{The background (\ref{background}) is
also time-dependent but we expect that particle production induced
by it is comparatively suppressed in the adiabatic regime $0 < \nu
\ll 1$.}. The Bogoliubov production amplitude is determined by
two-point correlators. MS computed the two-point correlator directly
(with their own prescriptions of analytic continuation and in
semiclassical approximation), and deduced that coherent Bogoliubov
pair production effectively shows a thermal distribution once
phase-correlations are averaged over.

Since the timelike theory is defined from the spacelike theory via
the Wick rotation (\ref{analy}), physical observables are
extractable accordingly. Indeed, taking the Wick rotation to the
reflection amplitude (\ref{D}) in Euclidean theory, we can extract
'reflection amplitude' in Lorentzian theory. This amplitude is
interpretable as the Bogoliubov coefficient of particle pair
production. We shall now show that the distribution functions
extracted so behaves in a manner anticipated by MS, but with
interesting and nontrivial twists due to full-fledged string
worldsheet effects. Our prescription (\ref{analy}) involves Wick
rotation of the Euclidean momentum to Lorentzian energy $a-q = ip
\to - \omega$, $\alpha \to i\kappa$ and $q \to iQ$. Taking also $Q
\rightarrow 0^+$,\footnote{This choice is automatically selected by
requiring that the ``Liouville wall" in the time-like theory is
located in the ``strongly coupled" region.} we obtain the Lorentzian
reflection amplitude ${\cal R}_{\rm L}$ as
\begin{eqnarray}
{\cal R}_{\rm L}(\omega,0,0) &=& (\xi_{\rm
L})^{-i\frac{\omega}{\kappa}} \, \frac{
\Gamma(1+i\frac{\omega}{\kappa})\Gamma(-i\frac{\omega} {\kappa})}
{\Gamma(1-i\frac{\omega}{\kappa}) \Gamma(+i\frac{\omega}{\kappa})} =
- (\xi_{\rm L})^{-i\frac{\omega}{\kappa}}
\ .
\label{reflo}
\end{eqnarray}
Here, $\xi_{\rm L}$ is the renormalized worldsheet cosmological
constant in the Lorentzian sine-Liouville theory, obtained from
(\ref{xi}) via the Wick rotation (\ref{analy}):
\bea \xi_{\rm L} = (\pi \mu)^2 {\gamma(-(-\kappa^2 + \beta^2 +
\delta^2)) \over \gamma(-\frac{1}{2}(-\kappa^2 + \beta^2 +
\delta^2))} \gamma(1 -  \kappa^2) \gamma(1 - \frac{1}{2}(\kappa^2 +
\beta^2 + \delta^2)) \ .\label{xiL} \eea
Here, the limit $\kappa^2 + \beta^2 + \delta^2 \rightarrow 2 +
\varepsilon$  ($\varepsilon >0$) is implicit (so that $Q = 0$ and
the string coupling parameter is set to constant).

Throughout the analytic continuation, the Bogoliubov coefficient
(\ref{reflo}) remains well-defined. Moreover, the first line in
(\ref{reflo}) is a pure phase except for the first factor involving
the renormalized cosmological constant $\xi_{\rm L}$. Now, suppose
$\xi_{\rm L}$ takes a negative value. In this case, there are two
possible vacuum branches, corresponding to the prescription
$\xi_{\rm L} = |{\xi}_{\rm L}| e^{\pm i \pi}$. We shall choose the
lower branch and obtain the Boltzmann-like, convergent particle
distribution function:
\begin{eqnarray}
{\cal P}(\omega) = \left| {\cal R}_{\rm L} \right|^2 \,\, = \,\,
\Big| \xi_{\rm L}^{-i\frac{\omega}{\kappa}}\Big|^2 =
e^{-\frac{2\omega}{T_{\rm eff}}}  \label{P omega} \ .
\end{eqnarray}
Here, the effective temperature $T_{\rm eff}$ is set entirely by the
Liouville coupling constant $\kappa$:
\begin{eqnarray}
T_{\rm eff} \equiv  {\kappa \over  \pi}.
\end{eqnarray}
This is precisely the sort of the behavior anticipated in \cite{MS},
except that the behavior here is deduced from entirely different
definition of the time-dependent tachyon background and from
different analytic continuations.

\subsection{phases of renormalized tachyon condensate}
According to (\ref{P omega}), the "Nothing state" --- state on which
no finite energy string excitation is possible --- arises in the
regime wherever the renormalized cosmological constant $\xi_{\rm L}$
gets negative-valued. We are thus interested in whether $\xi_{\rm
L}$ can change sign and, if it does, under what circumstances it
does.  We shall focus on $c=3$ system obtained by taking $Q
\rightarrow 0^+$ limit \footnote{If we take $Q \rightarrow 0^-$
limit, though we still obtain $c=3$ system, the "Nothing state"
obtained so have different characteristic. In particular, as we
shall discuss later, $Q \rightarrow 0^\pm$ theories exhibit quite
opposite behavior in the worldsheet semiclassical limit. This
indicates that there can be two distinct Lorentzian $c=3$ system
definable from the spacelike theory. We shall postpone such
non-analyticity for the moment, and return back to it later.}. Then,
$\xi_{\rm L}$ in (\ref{xiL}) takes the value:
\begin{eqnarray}
\xi_{\rm L} =
(\pi\mu)^2\frac{\gamma(-2(1-\kappa^2))}{\gamma(-(1-\kappa^2))}
\gamma(1-\kappa^2) \gamma(-0) \ . \label{xixi}
\end{eqnarray}
As it stands, $\xi_{\rm L}$ diverges because of the factor
$\gamma(-0)$. This divergence is multiplicative, so it is absorbable
by renormalizing the bare cosmological constant $\mu$
\footnote{Similar prescription was also proposed in the Lorentzian
Liouville theory \cite{stromingertakayanagi}.}. So, after the
renormalization, the sign of $\xi_{\rm L}$ depends solely on
$\kappa$, the Lorentzian Liouville exponent.

It is important to recall that the three-parameter sine-Liouville
theory has, in addition to the Liouville field $t$, compact boson
fields $\phi_1, \phi_2$. The parameter $\kappa$ controls interaction
of the Liouville field, while $\beta, \delta$ do so interaction of
the compact bosons $\phi_1, \phi_2$. In the previous subsection, we
took $Q \rightarrow 0^+$ so that we can set the string coupling
parameter $g_{\rm st}$ fixed to a small value.

\begin{figure}[htbp]
    \begin{center}
    \includegraphics[width=0.5\linewidth,keepaspectratio,clip]
      {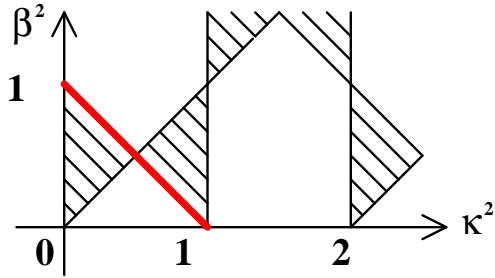}
    \end{center}
    \caption{\sl The sign of the renormalized cosmological constant
    $\xi_{\rm L}$ as a function of coupling parameters, $\kappa^2$
    and $\beta^2$. In the shaded/blank region, $\xi_{\rm L}$ takes
    positive/negative value. The solid red line corresponds to the
    $Q=0$ condition.}
    \label{sign}
\end{figure}

We are eventually interested in superstring case, for which we need
to fermionize $\phi_2$ by setting $\delta = 1$. The quantum
conformality condition (\ref{Lconfcond}) then relates $\phi_1$
dynamics coupling parameter $\beta$ to the Liouville coupling
parameter $\kappa$ as $\kappa^2 + \beta^2 = 1$. This implies that
the worldsheet dynamics is always strongly coupled: the $\phi_2$
field is kept strongly coupled by $\delta = 1$, while $t$ and
$\phi_1$ mutually interpolate between strong and weak coupling
regime by $\kappa^2 + \beta^2 = 1$. In other words, at least two out
of the three worldsheet fields $t, \phi_1, \phi_2$ are always
strongly coupled. In Fig.(\ref{sign}), we plot the regions in the
coupling parameter space, where the renormalized cosmological
constant $\xi_{\rm L}$ takes positive (blank) and negative (shaded)
values, respectively.
We also note that the condition $Q=0$ is a line located on the verge
of the sign change of $\xi_{\rm L}$. Thus, the parameter range we
are interested in indeed corresponds to strong coupling region. In
fact, exact conformal field theory approach beyond mini- or
midi-superspace approximation has been indispensable for us to
determine the actual sign of $\xi_{\rm L}$.

From the gamma functions, we see that $\xi_{\rm L} > 0$ for
$0<\kappa^2<\frac{1}{2}$ and $\xi_{\rm L} <0$ for
$\frac{1}{2}<\kappa^2<1$. Notice that, for superstring theories,
$\kappa^2 = 1$ sets the upper limit of the parameter space since the
limit corresponds to infinitely small compactification (or time-like
Liouville theory in the T-dual description). Accordingly, the
particle production probability behaves as
\begin{eqnarray}
{\cal P}(\omega) = \left| {\cal R}_{\rm L}(\omega) \right|^2 \,\,
\propto \,\, \left\{ \begin{array}{ccc} 1 & \qquad \mbox{for} &
\qquad 0 < \kappa^2 < {1 \over 2} \\ e^{-2\omega / T_{\rm eff}} &
\qquad \mbox{for} & \qquad \, \, {1 \over 2} < \kappa^2 < {1 }
 \ , \end{array}
\right.
\label{rl}
\end{eqnarray}
and the particle number distribution is determined to be
${\cal N}(\omega) = {\cal P}(\omega)/(1 \mp {\cal P}(\omega))$ for
boson and fermion particles, respectively.

Spectral moments of inclusive string production are then given by
\begin{eqnarray}
\langle E^n \rangle := \int_0^\infty \dd \omega \, \omega^n
\rho(\omega) {\cal N}(\omega) \qquad (n=0, 1, 2, \cdots)
\label{spectralmoments}
\end{eqnarray}
where $\rho(\omega)$ refers to density of closed string states. We
are eventually interested in our proposal in the superstring
context. In this case, the density of state behaves for large
$\omega$ as
\begin{eqnarray}
\rho(\omega) \sim e^{2 \pi \sqrt{c_{\rm eff} \over 3} \omega} ~
,
\end{eqnarray}
where $c_{\rm eff}$ is the `effective central charge' \cite{KutS}
of the transverse sector, which counts the net degrees of freedom.
In simple cases when the transverse
sector is a $D-2$ dimensional flat space with linear dilatons
$q_i$, the criticality condition is written as
\begin{eqnarray}
 \frac{3}{2}(D-2) + 3 \sum_i q_i^2 = 12~,
\end{eqnarray}
and we just have
\begin{eqnarray}
 c_{\rm eff} = 12- 3 \sum_i q_i^2 = \frac{3}{2}(D-2)~.
\label{c eff}
\end{eqnarray}
%

For $0<\kappa^2<\frac{1}{2}$, the transition probability
is flat, viz.
the effective temperature of the pair produced string distribution
is infinite. This may look
counter-intuitive since worldsheet dynamics of the Liouville field
$t$ becomes classical in the limit $\kappa \rightarrow 0$. However,
as discussed above, $Q=0$ and conformality conditions imply that ${1
\over 2} < \beta^2 < 1$ in this range. This means that the
$\phi_1$-field is strongly coupled. We thus interpret the divergent
string pair production as being triggered by strongly coupled
worldsheet dynamics of the compact boson fields, $\phi_1, \phi_2$.

For $\frac{1}{2}<\kappa^2<1$, the pair production probability is no
less than $e^{- 2 \sqrt{2} \pi \omega}$. The maximum is at $\kappa =
1$, at which the probability is $e^{- 2 \pi \omega}$. Now, the
density of string states $\rho(\omega)$ scales as $\rho(\omega) \sim
e^{2\pi \sqrt{c_{\rm eff}/3} \omega}$. Therefore, if $c_{\rm
eff}<3$, we always obtain ultraviolet finite particle production
irrespective of the value of $\kappa$. If $3<c_{\rm eff}<6$, we can
achieve ultraviolet finite particle production if the Liouville
coupling $\kappa$ is chosen suitably. If $c_{\rm eff}>6$, we always
face the ultraviolet divergence of the particle production for any
value of $\kappa$. In section 4, we will see that the threshold at
$c_{\rm eff} = 6$ corresponds to the string compactification on a
conifold and also curiously coincides with the "black hole / string
transition" \cite{nakayama1, amit, nakayama2} point $k=1$ of
SL$_k(2)/$U(1) coset conformal field theory. We note that, in this
range of Liouville coupling $\kappa$, worldsheet dynamics of the
Liouville field $t$ is strongly coupled, yet not strong enough to
warrant unsuppressed pair production as in the $0 < \kappa^2 < {1
\over 2}$ region. It implies that, in so far as the Bogoliubov pair
production is concerned, the compact bosons $\phi_1, \phi_2$ play
more significant role than the Liouville field $t$.

We thus conclude that, in the range ${1\over2} < \kappa^2 < 1$, the
Lorentzian three-parameter sine-Liouville theory (\ref{lorentzS})
yields an exactly soluble conformal field theory realization of the
"Nothing state" at the beginning of the Universe. In particular,
fermionic string theories with less than six dimensions (in the case
$D < 6$ in (\ref{c eff})) can have finite and controllable
Bogoliubov pair production and hence a viable model of nonsingular
string cosmology. We emphasize that the pair production behaves in
the present case quite differently from that in the Liouville
theory. As elaborated above, this is attributed to the
sine-Liouville potential involving $\phi_1, \phi_2$. We also
emphasize that, because we set $Q=0$ and the conformality condition,
we cannot achieve worldsheet dynamics entirely semiclassical: if the
Liouville field $t$ is weakly coupled the fields $\phi_1$ is
strongly coupled and vice versa.

\subsection{"Nothing state" from vacuum amplitude}
Alternative route for probing the "Nothing state" is to compute the
vacuum-to-vacuum transition amplitude and examine
whether the temporal
volume of the vacuum energy is cut off at a scale set by the Liouville
cosmological constant. Adopting the semiclassical approximation
\cite{wise}, MS estimated the one-loop
transition amplitude ${\cal Z}_{\rm L}$ as the product of
zero mode $Z_0$ and nonzero mode $\widehat{Z}$ contributions:
\begin{eqnarray}
{\cal Z}_{\rm L} = Z_0 \cdot \widehat{Z} \qquad \mbox{where} \qquad Z_0 =
\Big( - {1 \over \kappa} \ln {|\mu| \over \Lambda} - {i \over 2
T_{\rm eff}} \Big) \ ,
\end{eqnarray}
Here, $\ln \Lambda$ is the infrared cutoff for the Louville
direction. The real part (proportional to $\ln |\mu|$) is
interpreted as excising the temporal volume (out of the total volume
${ 1 \over \kappa} \ln \Lambda$) at the onset of "Nothing state".
The imaginary part is interpreted as exhibiting effective thermal
distribution of Bogoliubov pair production with temperature $T_{\rm
eff}$. In getting these results, however, it was technically crucial
for MS to rely on non-standard analytic continuation of the
cosmological constant $\mu \rightarrow i \mu$ they adopted.

From Lorentzian spacetime viewpoint, Bogoliubov pair production and
vacuum energy are related each other. The one-loop vacuum transition
amplitude can be understood as the time evolution under the
tree-level Bogoliubov pair production. Let us examine if such
relation can be checked directly by worldsheet analysis via the Wick
rotation (\ref{analy}). It is well-known that the reflection
amplitude ${\cal R}(p)$ of the Euclidean Liouville theory is related
to the density of states by
\begin{eqnarray}
\rho_{\rm finite}(p) = \frac{1}{2\pi i} \frac{{\rm d} {\cal R} (p)}
{{\rm d} p} \ .
\end{eqnarray}
Assuming that this assumption holds even after the Wick rotation to
the Lorentzian theory, we get
\begin{eqnarray}
\rho_{\rm finite}(\omega) = \frac{1}{2\pi i} \frac{ {\rm d} {\cal
R}_{\rm L}(\omega)}{{\rm d} \omega} \ . \label{dens}
\end{eqnarray}
We see that the Lorentzian relation (\ref{dens}) then leads to
appearance of the imaginary part, reaching the same conclusion as MS.
Let us now apply this general consideration to our proposal.
Substituting the two-point function \eqref{reflo} to \eqref{dens},
we obtain
\begin{eqnarray}
\rho_{\rm finite}(\omega) = - \frac{1}{2\pi  \kappa} \ln
\xi_{\rm L} + \rho_0(\omega) \  ,
\end{eqnarray}
where we separated explicitly the $\omega$-independent part from the
$\omega$-dependent part. The full density of states contains an
additional infrared cutoff factor $\ln \Lambda$. We are primarily
interested in the limit $\Lambda \rightarrow \infty$ (removing the
infrared cutoff). We then have
\begin{eqnarray}
{\cal Z}_{\rm torus} &=& \mathrm{Tr} \int \frac{\dd \omega}{2\pi} \,
e^{2\pi i\tau_1 P(\omega) -2\pi\tau_2 H(\omega)} \rho (\omega) \cr
    &\stackrel{\Lambda \to \infty}{\sim}&
    - \frac{1}{2 \kappa} \ln \left( \frac{\xi_{\rm L}}{\Lambda^2}\right)
    \int \frac{\dd\omega}{2\pi} \, \mathrm{Tr}
    \, e^{2\pi i\tau_1 P(\omega) -2\pi\tau_2 H(\omega)} \ ,
\end{eqnarray}
where $\mathrm{Tr}$ refers to summing over all other quantum numbers
than the energy.

In the previous section, we found that the {\it renormalized}
cosmological constant $\xi_{\rm L}$ is positive for $0 < \kappa^2 <
{1 \over 2}$ and becomes negative for $\frac{1}{2} <\kappa^2 < 1$.
The one-loop vacuum transition amplitude then takes the form:
\begin{eqnarray}
{\cal Z}_{\rm L} = \left\{ \begin{array}{cc}
\Big(-\frac{1}{2\kappa} \ln {|\xi_{\rm
L}|}/{\Lambda^2} -i0\Big) \widehat{Z} &
\qquad \mbox{for} \qquad  0 < \kappa^2 < {1 \over 2} \\
&\\
\Big( -\frac{1}{2\kappa} \ln{|\xi_{\rm L}|}/{\Lambda^2}
 - i \, \frac{\pi}{2\kappa} \Big) \widehat{Z} & \qquad \mbox{for} \qquad {1 \over 2}
 < \kappa^2 < 1. \end{array} \right.
\end{eqnarray}
We see that the result is in complete agreement with the tree-level
Bogoliubov amplitude result. The real part exhibits excision of the
time evolution of the universe at the beginning. Therefore, the
dependence of the one-loop vacuum amplitude on $-{1 \over 2\kappa}
\ln |\xi_{\rm L}|$ is a rigorous indication for the presence of the
"Nothing state". The imaginary part agrees perfectly with (\ref{rl})
in the real-time thermal field theory formalism: for two-particle
squeezed state, the time evolution runs over $[-\infty, t]$ and $[t,
t - i \beta_{\rm eff} / 2]$. We see from the above result that the
effective temperature $T_{\rm eff}$ is infinite for $0 < \kappa^2 <
{1 \over 2}$ and $\kappa/\pi$ for ${1 \over 2} < \kappa^2 < 1$,
respectively.

\subsection{Nothing state: from bare to renormalized}
The most striking feature of the "Nothing state" based on the
Lorentzian three-parameter sine-Liouville theory is the pattern of
the Bogoliubov particle production rate as given in (\ref{rl}).
Moreover, the rate (\ref{rl}) involves the renormalized cosmological
constant $\xi_{\rm L}$ and is the exact conformal field theory
result. Intuitively, we can interpret the result as follows. On
general grounds, we expect two possible sources that the
cosmological constant may change sign.
\begin{itemize}
    \item In the semiclassical, mini-superspace approximation
    of the Liouville theory, Wick rotation provides a source of
    the sign flip: the mini-superspace Hamiltonian
$ H_{\rm E} = + \partial_\varphi^2 + 4\pi \mu \, e^{\, \alpha \,
\varphi}$ turns under Wick rotation into $H_{\rm L} = -\partial_t^2
+ 4\pi \mu \, e^{-\kappa \, t}$. The Hamiltonian constraint in Euclidean
and Lorentzian cases are related each other by the flip of $\mu$ to
$-\mu$. Hence, the phase shift of the reflection amplitude $\sim
\mu^{ip}$ in the Euclidean theory becomes under Wick rotation the
damping factor of the Bogoliubov coefficient $\sim e^{-\omega \pi}$
in the Lorentzian theory.

In sine-Liouville theory, the situation is drastically different.
The Euclidean mini-superspace Hamiltonian
\begin{eqnarray}
H_{\rm E} = + \partial_\varphi^2 + \partial_{\phi_1}^2 +
\partial_{\phi_2}^2 + 2 \pi \mu e^{\alpha \varphi} \cos(\beta \phi_1
+ \delta \phi_2) \end{eqnarray}
becomes under the Wick rotation (\ref{analy}) the Lorentzian
mini-superspace Hamiltonian
\begin{eqnarray}
H_{\rm L} = - \partial_t^2 + \partial_{\phi_1}^2 +
\partial_{\phi_2}^2 + 2 \pi \mu e^{\alpha \varphi} \cos(\beta \phi_1
+ \delta \phi_2)
\end{eqnarray}
In both cases, the sign of bare cosmological constant is irrelevant
since it can be compensated by shifting $\beta \phi_1 + \delta
\phi_2$ by $\pi$. This explains why both the reflection amplitude
and the Bogoliubov coefficient depends on {\sl square} of $\mu$.
Thus, sign flip associated with the cosmological constant should
arise not from the bare one but from some other combination of the
coupling parameters.

    \item String worldsheet effects may provide
    another source of the
    sign flip. Take, for example, the ${\cal N}=2$
    ${\rm SL}_k (2, \mathbb{R})/{\rm U} (1)$ coset model describing
    the Euclidean two-dimensional black hole (cigar) geometry.
    The reflection amplitude behaves as
\begin{eqnarray}
{\cal R} (p) \, \, \propto \, \,
\left(M\frac{\Gamma(1-\frac{1}{k})}{\Gamma(1+\frac{1}{k})}
\right)^{ip} := \xi^{ip}, \label{sl2r}
\end{eqnarray}
where $M$ is the black hole mass, which plays via the FZZ duality
the role of {\sl bare} cosmological constant in ${\cal N}=2$
Liouville (sine-Liouville) theory. (See {\em e.g.} \cite{GK2}.)
We see from (\ref{sl2r}) that
what matters for the reflection amplitude is not the bare
cosmological constant $M$ but the renormalized one $\xi$. For
example, extrapolating the current algebra level $k$ naively across
1, we see that $\xi$ can change the sign.
\end{itemize}

\begin{figure}[htbp]
    \begin{center}
    \includegraphics[width=0.5\linewidth,keepaspectratio,clip]{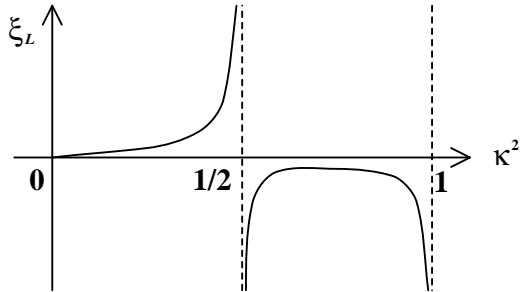}
    \end{center}
    \caption{\sl The renormalized cosmological constant $\xi_{\rm L}$ as
    a function of $\kappa^2$. We set $Q=0$ for an asymptotically flat
    universe. When
    $\xi_{\rm L}$ becomes negative as in the region $ \frac{1}{2} <\kappa^2
    < 1$, the Bogoliubov coefficient $\left|
    {\cal R}_{\rm L}(\omega) \right|$
    is exponentially suppressed.}
    \label{renorm}
\end{figure}

Based on these intuitions and considerations in the previous
subsection, we interpret the sign flip of $\xi_{\rm L}$ arising from
combination of both effects. For $0< \kappa^2 < \frac{1}{2}$, we
interpret $\xi_{\rm L} > 0 $ as a consequence of strong coupling
dynamics of the compact boson fields $\phi_1, \phi_2$. For ${1 \over
2} <\kappa^2<1$, we also interpret $\xi_{\rm L} <0$ as a consequence
of strong coupling dynamics of the Liouville field $t$. Behavior of
$\xi_{\rm L}$ as a function of $\kappa^2$ is plotted in
Fig.(\ref{renorm}).

\section{Embedding into Superstring Theories}
In the previous section, we proposed a viable string theory
realization of the "Nothing state". The proposal facilitates exactly
solvable conformal field theory approach, thus enabling us to define
the "Nothing state" precise enough. An immediate question is whether
our proposal can be made more realistic (i.e. free from bulk
tachyon) by embedding the model into superstring theories. In this
section, we shall argue that this can be done so {\sl provided} we
deform the sine-Liouville potential appropriately. In particular, we
shall show that the deformed fermionic theory admits superconformal
field theory description and features the same physics concerning
emergence of the "Nothing state" as the bosonic theory did.

One possible route for constructing exactly solvable cosmological
superstring background has been utilizing the Lorentzian Liouville
theory: the $c=1$ Liouville theory defined in terms of a timelike
worldsheet boson, free of screening charge. The Lorentzian Liouville
theory and tachyon condensation therein were much studied in recent
years \cite{gutperlestrominger,stromingertakayanagi}. In fact, the
MS proposal belongs to a variant of the Lorentzian Liouville theory.

In fact, there are several possible Lorentzian Liouville theories,
all classifiable according to the worldsheet supersymmetries.
Furthermore, they are all related to various limits of the
three-parameter sine-Liouville theory. Before presenting the
relevant model, in this section, we shall begin with features in
each of the models, with particular emphasis of advantage and
shortcomings in utilizing them for concrete realization of the MS
proposal.

\subsection{models with ${\cal N}=0$ conformal symmetry}
Begin with the bosonic (no worldsheet supersymmetry) Lorentzian
Liouville theory \cite{gutperlestrominger,stromingertakayanagi}.
Denoting the timelike worldsheet boson as $X^0$, the worldsheet
action~\footnote{Here, we set $\ell_{\rm st}^2 = 4\pi$.} of the
bosonic (no spacetime supersymmetry) string theory is given by
\bea S_{\rm TLT} = {1 \over 4 \pi} \int \dd^2 z \Big( - \partial X^0
\overline{\partial} X^0 + 4 \pi \mu \, e^{-2 \beta X^0} \Big) \eea
plus the action for flat $\mathbb{R}^{25}$ and for conformal ghosts.
With vanishing background charge, which amounts to constant-valued
dilaton, the worldsheet boson $X^0$ contributes central charge
$c=1$. The second term denotes to spatially homogeneous condensation
of the tachyon field: $T(X^0) = 4 \pi \mu \, \exp(-2 \beta X^0)$,
where the exponent $\beta$ is set to $\pm 1$ by the tachyon on-shell
condition. Choose the convention $\beta = +1$, viz. the tachyon
condensation grows exponentially at early epoch $X^0 \rightarrow
-\infty$. At late epoch $X^0 \rightarrow + \infty$, the tachyon
condensation is turned off, and the spacetime becomes flat
$\mathbb{R}^{1,25}$. Implicit to the consideration is that, for
real-valued $X^0$,  $\mu$ is restricted to positive definite value:
in the mini-superspace approximation,  the worldsheet
reparametrization invariance puts the constraint $(\dot{X}^0)^2 - 4
\pi \mu \, e^{-2 \beta X^0} = 0$.

As it stands, path integral of the Lorentzian Liouville theory is
ill-defined since the action is not positive definite. A
prescription proposed first by \cite{gutperlestrominger,
stromingertakayanagi} involves analytic continuation $(X^0, \beta)
\rightarrow (\phi, b) = (-i X^0, +i \beta)$. This is the standard
Wick rotation of the target spacetime. Taking the limit $b
\rightarrow \pm i$ along with the Wick rotation, the prescription
leaves the central charge $c=1$ intact \cite{fredenhagenschomerus}.
In the classical limit, mini-superspace analysis shows that
Bogoliubov pair production is exponentially suppressed. This agrees
with full-fledged worldsheet analysis, in which the renormalized
cosmological constant $\xi_{\rm L} = 2 \pi \mu \gamma(b^2)$ is found
to take negative value as $b \rightarrow \pm i 0$.

Schomerus \cite{schomerus} criticized the proposal of
Strominger-Takayanagi \cite{stromingertakayanagi}, and proposed
another prescription with results on two- and three-point
correlation functions. His proposal takes analytic continuation of
the central charge $c$: starting from the well-defined Liouville
theory with $c \ge 25$, two different limiting theories were
obtained in which conformal weights take real values. One is the
Euclidean Liouville theory and another is the Lorentzian Liouville
theory. The resulting two theories are not continuable by the Wick
rotation proposed in \cite{stromingertakayanagi}, and this also
explains nonanalytic factors in three-point correlation functions.
In so far as two-point correlation functions are concerned, both
prescriptions turn out to yield the same result. Moreover, within
the mini-superspace approximation, \cite{fredenhagenschomerus2}
argued that the Bogoliubov amplitude is always unitary for a given
choice of self-adjoint Liouville Hamiltonian. It was also observed
that, if summed over all possible choice of self-adjointness, the
Bogoliubov amplitude becomes exponentially damped and coincides with
the result of \cite{stromingertakayanagi}.

The most serious difficulty of the bosonic Liouville theory is that,
being bosonic string theory, the bulk tachyon is present in the
spectrum: even at late epoch when the winding string tachyon
background is turned off, the bulk tachyon would destabilize
the flat spacetime $\mathbb{R}^{1,25}$.

\subsection{models with ${\cal N}=1$ superconformal symmetry}
The Lorentzian Liouville theory with ${\cal N}=1$ worldsheet
supersymmetry is the simplest generalization of the ${\cal N}=0$
bosonic counterpart. With the ${\cal N}=1$ worldsheet supersymmetry
alone, spacetime supercharges are not constructible. So, this class
of models describes tachyon condensation of Type 0 string theories.
The theory may be defined via the standard Wick rotation of the
${\cal N}=1$ Euclidean Liouville theory, whose worldsheet action is
given in terms of the ${\cal N}=1$ supermultiplet $(\varphi, \psi,
\widetilde{\psi})$ by
\bea S_{{\cal N}=1} = {1 \over 2 \pi} \int \dd^2 z \,\Big( \partial
\varphi \overline{\partial} \varphi + i \psi \overline{\partial}
\psi + i \widetilde{\psi} \partial \widetilde{\psi} + {1 \over 4} Q
\varphi {\cal R}_{(2)} + 2 \pi \mu b^2\,  \widetilde{\psi} \psi e^{b
\varphi} \Big) \ , \eea
where $Q = (b + 1/b)$. In fact, the ${\cal N}=1$ superconformal
Liouville potential represents tachyon background with no winding
mode excitations. In other words, the closed string tachyon is the
bulk tachyon.

Furthermore, in this model, the Bogoliubov pair production is
unsuppressed, as can be derived both from mini-superspace
approximation and from exact two-point function. In fact, in
manifestly ${\cal N}=1$ supersymmetric formulation, the
mini-superspace approximation is inert to the Wick rotation. This is
because the ${\cal N}=1$ supersymmetric Hamiltonian constraint is
given as (in NS-NS sector)
\begin{eqnarray}
\left[ - \Big({\partial \over \partial \varphi}\Big)^2 +
\Big({\partial W \over \partial \varphi}\Big)^2 \right]
\Psi(\varphi) = 0 \ ,
\end{eqnarray}
where $W$ is the superpotential, and it changes only by an {\em
overall} sign under the Wick rotation $\varphi \rightarrow -i X^0$.
This also fits with the conformal field theory analysis, in which
the renormalized cosmological constant is given by $\xi_{\rm L} = 2
\pi \mu \gamma (bQ/2)$ \cite{rashkov, Ahn} and takes positive value
as $b \rightarrow \pm i 0$. Here, we are following the conventions
of \cite{Fukuda:2002bv} \cite{Nakayama:2004vk}.

\subsection{models with ${\cal N}=2$ superconformal symmetry}
One can also start with Euclidean Liouville theory with ${\cal N}=2$
worldsheet superconformal invariance and prescribe a Lorentzian
counterpart by Wick rotation. Closely related model (`T-dual
version') is constructed by the formal analytic continuation
$k\,\rightarrow\, -k$ in the SL$_k(2, \mathbb{R})$/U(1)
(super)coset, which is often called as the `cosmological' SL$_k(2,
\mathbb{R})$/U(1) model \cite{CKR,HT,TT}.
These theories are easily
embeddable into Type II superstring vacua with
unbroken space-time supersymmetry.
There is potentially a tachyonic infrared
instability due to the negative mass gap along the timelike linear
dilaton direction \cite{CKR}, but this is easily removable by
adding spacelike linear dilaton (as in the light-like linear
dilaton models \cite{CSV,Craps}).

However, these models also have a serious difficulty of ultraviolet
divergence due to particle production. The particle production rate
behaves \cite{HT,TT} for small radial momenta
as $\sim{\cal O}(1)$ for all frequencies since there is no
exponential damping. We thus face a
uncontrollable ultraviolet divergence caused by the Hagedorn
density of states.
One may attempt to avoid this difficulty by taking the level $k$ of
SL$_k(2, \mathbb{R})$/U(1) coset conformal field theory to $|k|<1$ (see
discussions around \eqref{sl2r}).  However, in the cosmological
model, $|k|<1$ means that the central charge is negative,
thus it does not offer a sensible resolution.

A further difficulty has to do with `wrong' sign Liouville wall:
unlike ${\cal N} \le 1$ counterparts, the sine-Liouville potential
in this case is peaked at weak coupling region. This is inevitable
because of the condition $2 \alpha q = 1$, which is need if the
${\cal N}=2$ worldsheet superconformal symmetry is to be retained
\footnote{In \cite{HT}, based on an interpretation of this
peculiarity, a possibility was suggested  for removing the
cosmological singularity at string theory level. However, the MS
(with `correct' Liouville wall) seems more plausible.}. The
resulting theory is known as the two-parameter sine-Liouville model
\cite{fateev}.
This is because the worldsheet action
\begin{eqnarray}
S = \frac{1}{2\pi}\int \dd^2 z \Big[ \partial\varphi \bar{\partial}\varphi +
\partial\phi_1\bar{\partial}\phi_1 + \partial\phi_2\bar{\partial}\phi_2
+ 2\pi \mu \, e^{\alpha\varphi}\cos(\beta\phi_1+ \delta\phi_2) + \frac{1}{4\alpha} \varphi {\cal R}_{(2)} \Big] \nonumber
\end{eqnarray}
is changed to
\begin{eqnarray}
S = \frac{1}{2\pi}\int \dd^2 \sigma \Big[ -\partial t\bar{\partial}t +
\partial\phi_1\bar{\partial}\phi_1 + \partial\phi_2\bar{\partial}\phi_2
+ 2\pi \mu \, e^{-\kappa t}\cos(\beta\phi_1+ \delta\phi_2)
+ \frac{1}{4\kappa} t {\cal R}_{(2)} \Big] \nonumber
\end{eqnarray}
after the Wick rotation of $\alpha \to i \kappa$ and
$\varphi \to it$. The resulting Lorentzian theory is pathological
since the strong coupling region is not shielded by the
Liouville potential.
Although the conformal conditions are satisfied,
the theory has a strong coupling singularity at $ t \to +\infty$.
The difficulty arises because of
{\it wrong} Liouville potential term $\propto e^{- \kappa t}$,
which is peaked at $t \rightarrow - \infty$ and decreases as
$t\to + \infty$.
(Compare this with the worldsheet action of three-parameter theory
(\ref{lorentzS}).)
This difficulty, which does not exist in
the bosonic Liouville theory~\footnote{This is essentially due to the
fact that the background charge is given by $q= b+1/b$ in bosonic
Liouville theory and hence allows more choices of the sign after
Lorentzian continuation $b \to i\beta$.}, has led us to consider the
three-parameter version in this work.

\subsection{three-parameter sine-Liouville theory:\\
deformation to ${\cal N} = 1$ superconformal model}
Finally, let us try to embed the three-parameter sine-Liouville
model, studied in section 3,
into superstring theory.
This would lead to the most sensible cosmological model
because of the following reasons:
\begin{enumerate}
 \item We do not have the bulk closed string tachyon causing
 the infrared instability in the similar manner as
 the ${\cal N}=2$ models
 (or the cosmological $\mbox{SL}(2, \mathbb{R})/\mbox{U}(1)$ models)
 as far as embedding
 into the GSO projected superstring backgrounds.
 \item The particle production rate could be exponentially
small exceeding the Hagedorn growth of state density
similarly to the ${\cal N}=0$ case,
if the parameters are chosen suitably. This means that we are
also free from the ultraviolet instability.
 \item We have the parameter region in which the Liouville wall
is located at the `correct' side, which makes the quantum
theory well-defined. This fact sharply contrasts with
the ${\cal N}=2$ models .
\end{enumerate}

However, unfortunately
the three-parameter model does not possess any
worldsheet supersymmetry even if the scalar field $\phi_2$
is fermionized (except the ${\cal N}=2$
enhancement for the two-parameter model).
Instead, we may consider an alternative
model with ${\cal N} =1$ supersymmetric interaction
\begin{eqnarray}
S_{\rm int} = \int \dd^2 z \dd^2\theta \, \mu e^{\alpha \Xi}\, \cos(\beta \Phi_1)  \label{n1model}
\end{eqnarray}
where $\Xi, \Phi_1$ are ${\cal N}=1$ superfields with components
$(\varphi, \psi, \widetilde{\psi})$ and
$(\phi_1, \psi_1, \widetilde{\psi}_1)$, respectively. Fortuitously, this ${\cal N}=1$ model
bears exactly the same structure of the two-point correlators in the {\it neutral sector} as the three-parameter models discussed in section 3  except trivial rescaling of the bare cosmological constant:
\begin{eqnarray}
\mu^2 \to (\alpha^2 +\beta^2)^2 \mu^2 \ . \label{rescaling}
\end{eqnarray}
This is because the only modification in the derivation of the
reflection amplitude and the Bogoliubov amplitude is to replace
the screening insertions
\begin{eqnarray}
\mu^2 \int \dd^2z_1 \dd^2z_2 \, : e^{\alpha\varphi + i\beta\phi_1
+i\delta \phi_2}:(z_1) \, :e^{\alpha\varphi-i\beta\phi_1
-i\delta \phi_2}:(z_2)
\end{eqnarray}
in (\ref{3parasl})  by
\begin{eqnarray}
\mu^2  \int \dd^2z_1 \dd^2z_2 \,
:(\alpha^2 \psi \widetilde{\psi}+i \alpha\beta \psi\widetilde{\psi}_1
+ i \alpha\beta \psi_1\widetilde{\psi} - \beta^2\psi_1\widetilde{\psi}_1)
e^{\alpha\varphi+i\beta\phi_1}:(z_1)
\cr
\times:(\alpha^2 \psi\widetilde{\psi}-i \alpha\beta \psi\widetilde{\psi}_1
-i \alpha\beta \psi_1\widetilde{\psi} - \beta^2\psi_1\widetilde{\psi}_1)e^{
\alpha\varphi-i \beta\phi_1}: (z_2) & &\ .
\end{eqnarray}
The two-point correlators in the neutral sector do not involve the
fermions. So, contracting them among themselves, one effectively has
screening operators exactly the same as those of the three-parameter
model except rescaling the coupling parameters as in
(\ref{rescaling}) .

As such, the ${\cal N}=1$ model with the worldsheet interaction
(\ref{n1model}) would provide a realistic "Nothing state" in
the context of fermionic string theories.
Since the Bogoliubov particle production would be dominated by the
neutral sector, we can adopt the reflection amplitudes of the
three-parameter sine-Liouville theory extracted in section 3
to the present context.

In the previous section, we showed that a technically natural string
theory setup of "Nothing state" can be realized if the effective
central charge $c_{\rm eff}$ is less than or equal to 6. What kind
of background does it correspond to? To answer this, let us consider
a string theory background of the type
\begin{eqnarray}
\mathbb{R}^2_\perp \times \mathbb{R}_{Q} \times {\cal M} \ ,
\end{eqnarray}
where $\mathbb{R}^2_\perp$ is identified as the transverse part of
the four-dimensional spacetime, $\mathbb{R}_Q$ as ${\cal N}=2$
Liouville theory with dilaton slope $Q$, and ${\cal M}$ as a unitary
conformal field theory. Then, by the conformality condition $c_{\rm
tot}$=12, we find that the critical situation $c_{\rm eff}=6$ is
attained if $Q$ equals to $\sqrt{2}$ and $c_{\cal M}=0$ (i.e. ${\cal
M}$ sector should be trivial). This situation is quite interesting
since it is interpretable as the four-dimensional superstring
compactified on a conifold! The relation between $\mathbb{R}_Q$ and
the conifold is well-known \cite{Ghoshal:1995wm, Giveon:1999zm}. It
is quite interesting that conifolds show up prominently in
constructing nonsingular string cosmology. In this context, recall
that most Calabi-Yau threefolds have conifold points in their moduli
spaces and that the density of supersymmetric flux vacua is sharply
peaked near the conifold points~\cite{douglas, eguchi}. This may be
an indication that nonsingular string cosmology model proposed in
this work is abundantly realizable out of supersymmetric flux vacua.

The critical coupling $Q = \sqrt{2}$ coincides via FZZ duality with
the critical level $k=1$ of SL$_k(2, \mathbb{R})$ supercoset.
Curiously, this is precisely where the "black hole / string
transition" is known to take place \cite{nakayama1, amit, nakayama2}
for string theory on three-dimensional anti-de Sitter (AdS$_3$)
background with curvature $\sqrt{k} \ell_{\rm st}$ and for string
theory on linear dilaton background with dilaton slope $Q =
\sqrt{2/k}$. The phase $c_{\rm eff} < 6$ for realizing the "Nothing
state" coincides with the phase of long excited strings either at
boundary of the AdS$_3$ or throat of the linear dilaton background.

Technically, the coincidence has to do with growth of Hagedorn
density of states and ultraviolet behavior therein. For a D-brane
rolling in five-brane \cite{reyfivebrane, CHS} and related
backgrounds, we found in \cite{nakayama1, nakayama2} that the closed
string emission is ultraviolet finite if $k \le 1$ but divergent for
$k > 1$. For the winding tachyon condensation, we found above that
the Bogoliubov pair production is ultraviolet finite if $k \le 1$
but divergent for $k >1$. Whether the coincidence bears deeper
connection between the two situations poses a very interesting
question. We intend to report progress on this in a separate work.

\subsection{intuitive spacetime picture}
We end the discussion with a heuristic remark that may explain how
the "Nothing state" can be understood intuitively. The crux of MS
proposal is that winding string becomes tachyonic in an epoch near
the cosmological singularity. Let us treat for simplicity the
tachyon mass a constant-valued throughout the entire spacetime.
Then, the effective field theory of the tachyon field $T$ coupled to
the gravity $g_{mn}$ is given by
\bea S_{10d} = \int d^{10} x \sqrt{-g} \, \Big[ {\cal R} + {1 \over
2} (\nabla_m T)^2 + {\kappa^2 \over 2} T^2 + \cdots \Big]. \eea
Here, $-\kappa^2 <0 $ denotes the mass-squared of the winding string
tachyon field. Taking the ansatz of the Einstein-de Sitter space for
the metric:
\bea \dd s^2 = - \dd t^2 + a^2(t) d {\bf x}^2 \qquad \mbox{and}
\qquad T(x) = T(t) \eea
the field equations read
\bea && \left({\dot{a} \over a}\right)^2 = {1 \over 6} (\dot{T}^2 -
\kappa^2 T^2) \nonumber \\
&& \ddot{T} + 9 {\dot{a} \over a} \dot{T} - \kappa^2 T = 0. \eea
An obvious solution is that $T(t) = 0$ and $a(t)$ takes a constant
value. This describes a static universe in which the tachyon is not
condensed. Another solution is that $T(t) = \exp ( - \kappa t)$ and
$a(t)$ takes again a constant value. This describes again a static
and nonsingular universe in which the tachyon is coherently
condensed. For both cases, the universe is
static since the energy vanishes identically. Obviously,
the latter solution approaches to the first
as $t \rightarrow +\infty$. In general, the tachyon may
evolve differently and cause the spacetime to evolve cosmologically.
Thus, starting from the solution $T(t) = \exp (-\kappa t)$ as the
asymptotic solution in the infinite past $t \rightarrow - \infty$,
by continuity, one can interpolate to the solution $T(t) = 0$ via
the cosmological solution. It then leads to the beginning of
cosmological universe starting from a nonsingular, static universe.

\section*{ Acknowledgement}

We thank Changrim Ahn, Dan Israel, Hirosi Ooguri, Sylvain Ribault,
Volker Schomerus, Steve Shenker and Tadashi Takayanagi for useful
discussions. We acknowledge that part of this work was carried out
while participating in the focus program "Liouville, D-branes and
Integrability" at the Asia Pacific Center for Theoretical Physics
(APCTP). This work was supported in part by the JSPS Research
Fellowship for Young Scientists (YN), Ministry of Education,
Culture, Sports, Science and Technology of Japan (YS),
the KOSEF SRC Program through "Center for Quantum Space-Time"
(R11-2005-021) (SJR).
%


\begin{thebibliography}{99}

\bibitem{martinec}
  A.~E.~Lawrence and E.~J.~Martinec,
  Class.\ Quant.\ Grav.\  {\bf 13}, 63 (1996)
  [arXiv:hep-th/9509149].

\bibitem{rey}
S.~J.~Rey,
  Phys.\ Rev.\ Lett.\  {\bf 77}, 1929 (1996)
  [arXiv:hep-th/9605176];
  S.~J.~Rey,
  Nucl.\ Phys.\ Proc.\ Suppl.\  {\bf 52A}, 344 (1997)
  [arXiv:hep-th/9607148];
S.~J.~Rey,
  arXiv:hep-th/9609115.


  \bibitem{larsenwilczek}
  F.~Larsen and F.~Wilczek,
  Phys.\ Rev.\ D {\bf 55}, 4591 (1997)
  [arXiv:hep-th/9610252].

\bibitem{Balasubramanian:2002ry}
  V.~Balasubramanian, S.~F.~Hassan, E.~Keski-Vakkuri and A.~Naqvi,
  Phys.\ Rev.\ D {\bf 67}, 026003 (2003)
  [arXiv:hep-th/0202187].

\bibitem{cornalba}
  L.~Cornalba, M.~S.~Costa and C.~Kounnas,
  Nucl.\ Phys.\ B {\bf 637}, 378 (2002)
  [arXiv:hep-th/0204261].

\bibitem{Gasperini:2002bn}
  M.~Gasperini and G.~Veneziano,
  Phys.\ Rept.\  {\bf 373}, 1 (2003)
  [arXiv:hep-th/0207130].

\bibitem{easson}
D.~A.~Easson,
  Phys.\ Rev.\ D {\bf 68}, 043514 (2003)
  [arXiv:hep-th/0304168].


\bibitem{MS}
  J.~McGreevy and E.~Silverstein,
  arXiv:hep-th/0506130.




\bibitem{polyakov}
  A.~M.~Polyakov,
  ``A Few projects in string theory,''
  arXiv:hep-th/9304146.



\bibitem{atickwitten}
J.J. Atick and E. Witten,
Nucl. Phys. B \bf 310 \rm, 291 (1988).


\bibitem{scherkschwarz}
  J.~Scherk and J.~H.~Schwarz,
  Phys.\ Lett.\ B {\bf 82}, 60 (1979).





\bibitem{GSO}
  F.~Gliozzi, J.~Scherk and D.~I.~Olive,
  Nucl.\ Phys.\ B {\bf 122}, 253 (1977).


\bibitem{BarRab}
J.L.F. Barbon and E. Rabinovici,
arXiv:hep-th/0407236.

\bibitem{ALMSS}
  A.~Adams, X.~Liu, J.~McGreevy, A.~Saltman and E.~Silverstein,
  JHEP {\bf 0510}, 033 (2005)
  [arXiv:hep-th/0502021].



\bibitem{Horo}
  G.~T.~Horowitz,
  JHEP {\bf 0508}, 091 (2005)
  [arXiv:hep-th/0506166].

\bibitem{Ross}
  S.~F.~Ross,
  JHEP {\bf 0510}, 112 (2005)
  [arXiv:hep-th/0509066].

\bibitem{BerHir}
  O.~Bergman and S.~Hirano,
  Nucl.\ Phys.\ B {\bf 744}, 136 (2006)
  [arXiv:hep-th/0510076].

\bibitem{IKS}
  N.~Itzhaki, D.~Kutasov and N.~Seiberg,
  JHEP {\bf 0512}, 035 (2005)
  [arXiv:hep-th/0510087].

\bibitem{HorSil}
  G.~T.~Horowitz and E.~Silverstein,
  Phys.\ Rev.\ D {\bf 73}, 064016 (2006)
  [arXiv:hep-th/0601032].


\bibitem{BerkPioRoz}
  M.~Berkooz, B.~Pioline and M.~Rozali,
  JCAP {\bf 0408}, 004 (2004)
  [arXiv:hep-th/0405126].

\bibitem{HikTai}
  Y.~Hikida and T.~S.~Tai,
  JHEP {\bf 0601}, 054 (2006)
  [arXiv:hep-th/0510129].


\bibitem{EGKR}
  S.~Elitzur, A.~Giveon, D.~Kutasov and E.~Rabinovici,
  JHEP {\bf 0206}, 017 (2002)
  [arXiv:hep-th/0204189];

\bibitem{CKR}
  B.~Craps, D.~Kutasov and G.~Rajesh,
  JHEP {\bf 0206}, 053 (2002)
  [arXiv:hep-th/0205101];

\bibitem{BCKR}
  M.~Berkooz, B.~Craps, D.~Kutasov and G.~Rajesh,
  JHEP {\bf 0303}, 031 (2003)
  [arXiv:hep-th/0212215].


\bibitem{HT}
  Y.~Hikida and T.~Takayanagi,
  Phys.\ Rev.\ D {\bf 70}, 126013 (2004)
  [arXiv:hep-th/0408124].

\bibitem{TT}
  N.~Toumbas and J.~Troost,
  JHEP {\bf 0411}, 032 (2004)
  [arXiv:hep-th/0410007].


\bibitem{Craps}
  B.~Craps,
  arXiv:hep-th/0605199.









\bibitem{FZZ2}
V.~Fateev, A.~B.~Zamolodchikov and Al.~B.~Zamolodchikov,
unpublished;
  V.~Kazakov, I.~K.~Kostov and D.~Kutasov,
  Nucl.\ Phys.\ B {\bf 622}, 141 (2002)
  [arXiv:hep-th/0101011].


\bibitem{fateev}
  P.~Baseilhac and V.~A.~Fateev,
  Nucl.\ Phys.\ B {\bf 532}, 567 (1998)
  [arXiv:hep-th/9906010].

\bibitem{Kim}
  J.~Kim, B.~H.~Lee, C.~Park and C.~Rim,
  J. Korean Phys. Soc. {\bf 46}, 1311 (2005)
  [arXiv:hep-th/0503050].


\bibitem{hartlehawking}
  J.~B.~Hartle and S.~W.~Hawking,
  Phys.\ Rev.\ D {\bf 28}, 2960 (1983).

\bibitem{FZZ}
V.~Fateev, A.~B.~Zamolodchikov and A.~B.~Zamolodchikov,
  arXiv:hep-th/0001012.






\bibitem{GK}
A.~Giveon and D.~Kutasov,
JHEP {\bf 9910}, 034 (1999)
[arXiv:hep-th/9909110];
A.~Giveon and D.~Kutasov,
JHEP {\bf 0001}, 023 (2000)
[arXiv:hep-th/9911039].



\bibitem{HoriKap}
K.~Hori and A.~Kapustin,
JHEP {\bf 0108}, 045 (2001)
[arXiv:hep-th/0104202].


\bibitem{Tong}
D.~Tong,
JHEP {\bf 0304}, 031 (2003)
[arXiv:hep-th/0303151].



\bibitem{Zamolodchikov:1995aa}
  A.~B.~Zamolodchikov and A.~B.~Zamolodchikov,
  Nucl.\ Phys.\ B {\bf 477}, 577 (1996)
  [arXiv:hep-th/9506136].





\bibitem{gutperlestrominger}
  M.~Gutperle and A.~Strominger,
  Phys.\ Rev.\ D {\bf 67}, 126002 (2003)
  [arXiv:hep-th/0301038].

\bibitem{stromingertakayanagi}
  A.~Strominger and T.~Takayanagi,
  Adv.\ Theor.\ Math.\ Phys.\  {\bf 7}, 369 (2003)
  [arXiv:hep-th/0303221].


\bibitem{wise}
A. Gupta, S.P. Trivedi and M.B. Wise,
Nucl.\ Phys.\ B {\bf 340}, 475 (1990);\\
M. Bershadsky and I.R. Klebanov,
Phys. \ Rev. \ Lett. \ {\bf 65}, 3088 (1990).




\bibitem{KutS}
D.~Kutasov and N.~Seiberg,
Nucl.\ Phys.\ B {\bf 358}, 600 (1991).


\bibitem{nakayama1}
Y.~Nakayama, K.~L.~Panigrahi, S.~J.~Rey and H.~Takayanagi,
  JHEP {\bf 0501}, 052 (2005)
  [arXiv:hep-th/0412038].

\bibitem{amit}
A.~Giveon, D.~Kutasov, E.~Rabinovici and A.~Sever,
  Nucl.\ Phys.\ B {\bf 719}, 3 (2005)
  [arXiv:hep-th/0503121].

\bibitem{nakayama2}
 Y.~Nakayama, S.~J.~Rey and Y.~Sugawara,
  JHEP {\bf 0509}, 020 (2005)
  [arXiv:hep-th/0507040].


\bibitem{reyfivebrane}
S.~J.~Rey,
  Phys.\ Rev.\ D {\bf 43}, 526 (1991);
ibid. "{\sl Axionic String Instantons and Their Low-Energy
Implications}", UCSB-TH-89/49, in the proceedings of the "Workshop
on Superstrings and Particle Theory" pp. 291-300 (World Scientific
Pub., 1989).

\bibitem{CHS}
C.~G.~.~Callan, J.~A.~Harvey and A.~Strominger,
Nucl.\ Phys.\ B {\bf 359}, 611 (1991);
S.~J. Rey,  "{\sl On String Theory and Axionic Strings and
Instantons}", SLAC-PUB-5659, in the proceedings of "Particle and
Fields '91 Conference" pp. 876-881 (American Physical Society,
1991).


\bibitem{GK2}
  A.~Giveon and D.~Kutasov,
  Nucl.\ Phys.\ B {\bf 621}, 303 (2002)
  [arXiv:hep-th/0106004].





\bibitem{fredenhagenschomerus}
S.~Fredenhagen and V.~Schomerus,
  JHEP {\bf 0505}, 025 (2005)
  [arXiv:hep-th/0409256].


\bibitem{schomerus}
 V.~Schomerus,
  JHEP {\bf 0311}, 043 (2003)
  [arXiv:hep-th/0306026].




\bibitem{fredenhagenschomerus2}
  S.~Fredenhagen and V.~Schomerus,
  JHEP {\bf 0312}, 003 (2003)
  [arXiv:hep-th/0308205].



\bibitem{rashkov}
  R.~C.~Rashkov and M.~Stanishkov,
  Phys.\ Lett.\ B {\bf 380}, 49 (1996)
  [arXiv:hep-th/9602148];
  See also R.~H.~Poghosian,
  Nucl.\ Phys.\ B {\bf 496}, 451 (1997)
  [arXiv:hep-th/9607120].

\bibitem{Ahn}
  C.~Ahn, C.~Rim and M.~Stanishkov,
  Nucl.\ Phys.\ B {\bf 636}, 497 (2002)
  [arXiv:hep-th/0202043].



\bibitem{Fukuda:2002bv}
  T.~Fukuda and K.~Hosomichi,
  Nucl.\ Phys.\ B {\bf 635}, 215 (2002)
  [arXiv:hep-th/0202032].
\bibitem{Nakayama:2004vk}
  Y.~Nakayama,
  Int.\ J.\ Mod.\ Phys.\ A {\bf 19}, 2771 (2004)
  [arXiv:hep-th/0402009].





\bibitem{CSV}
  B.~Craps, S.~Sethi and E.~P.~Verlinde,
  arXiv:hep-th/0506180.

\bibitem{Ghoshal:1995wm}
  D.~Ghoshal and C.~Vafa,
  Nucl.\ Phys.\ B {\bf 453}, 121 (1995)
  [arXiv:hep-th/9506122].

\bibitem{Giveon:1999zm}
  A.~Giveon, D.~Kutasov and O.~Pelc,
  JHEP {\bf 9910}, 035 (1999)
  [arXiv:hep-th/9907178].


\bibitem{douglas}
S.~Ashok and M.~R.~Douglas,
  JHEP {\bf 0401}, 060 (2004)
  [arXiv:hep-th/0307049];
F.~Denef and M.~R.~Douglas,
  JHEP {\bf 0405}, 072 (2004)
  [arXiv:hep-th/0404116].


\bibitem{eguchi}
T.~Eguchi and Y.~Tachikawa,
  JHEP {\bf 0601}, 100 (2006)
  [arXiv:hep-th/0510061].



\end{thebibliography}
\end{document}